\documentstyle[epsf,12pt]{article}

\catcode`\@=11

\def\prefixsection{\S}

\ifx\documentclass\undefined
 \def\@sect#1#2#3#4#5#6[#7]#8{\ifnum #2>\c@secnumdepth
     \let\@svsec\@empty\else
     \refstepcounter{#1}\edef\@svsec{\csname prefix#1\endcsname
	\csname the#1\endcsname\hskip 1em}\fi
     \@tempskipa #5\relax
      \ifdim \@tempskipa>\z@
        \begingroup #6\relax
          \@hangfrom{\hskip #3\relax\@svsec}{\interlinepenalty \@M #8\par}%
        \endgroup
       \csname #1mark\endcsname{#7}\addcontentsline
         {toc}{#1}{\ifnum #2>\c@secnumdepth \else
                      \protect\numberline{\csname the#1\endcsname}\fi
                    #7}\else
        \def\@svsechd{#6\hskip #3\relax  
                   \@svsec #8\csname #1mark\endcsname
                      {#7}\addcontentsline
                           {toc}{#1}{\ifnum #2>\c@secnumdepth \else
                             \protect\numberline{\csname the#1\endcsname}\fi
                       #7}}\fi
     \@xsect{#5}}
\else
    \def\@seccntformat#1{\csname prefix#1\endcsname
	\csname the#1\endcsname\quad}
\fi

\@addtoreset{equation}{section}   
\def\theequation{\arabic{section}.\arabic{equation}}

\def\thebibliography#1{\section*{References\@mkboth
 {REFERENCES}{REFERENCES}}\list
 {\leftbibmark\arabic{enumi}\rightbibmark}{
 \settowidth\labelwidth{\leftbibmark #1\rightbibmark}\leftmargin\labelwidth
 \advance\leftmargin\labelsep
 \usecounter{enumi}}
 \def\newblock{\hskip .11em plus .33em minus -.07em}
 \sloppy\clubpenalty4000\widowpenalty4000
 \sfcode`\.=1000\relax}

\def\@citex[#1]#2{\if@filesw\immediate\write\@auxout{\string\citation{#2}}\fi
  \def\@citea{}\@cite{\@for\@citeb:=#2\do
    {\@citea\def\@citea{,\penalty\@m\ }\@ifundefined
       {b@\@citeb}{{\bf ?}\@warning
       {Citation `\@citeb' on page \thepage \space undefined}}%
\hbox{\csname b@\@citeb\endcsname\citemarkdelim}}}{#1}}

\def\@cite#1#2{\leftcitemark{#1 \if@tempswa , #2\fi}\rightcitemark}
\def\leftcitemark{[}
\def\rightcitemark{]}
\def\citemarkdelim{}
\def\leftbibmark{[}
\def\rightbibmark{]}

\catcode`\@=12

\def\D{{\cal D}}

\def\H{{\cal H}}

\def\L{{\cal L}}

\def\Eq#1{Eq.(\ref{#1})}

\def\r#1{{\rm #1}}

\def\bm#1{\mbox{\boldmath $#1$}}
\let\bg=\bm
\def\Frac(#1/#2){\left(\frac{#1}{#2}\right)}
\def\Tr{\r{Tr}}

\def\gsim{\stackrel{>}{\sim}}

\def\Order#1{\r{O}\!\left(#1\right)}

\def\Beq{\begin{equation}}
\def\Eeq{\end{equation}}
\def\Beqr{\begin{eqnarray}}
\def\Eeqr{\end{eqnarray}}
\def\Beqrn{\begin{eqnarray*}}
\def\Eeqrn{\end{eqnarray*}}
\def\Bitm{\begin{itemize}}
\def\Eitm{\end{itemize}}

\font\elevenmib=cmmib10 scaled\magstephalf   \skewchar\elevenmib='177
\def\YUKAWAmark{\hbox{\elevenmib 
 Yukawa\hskip0.05cm Institute\hskip0.05cm Kyoto \hfill}}

\begin{document}
\def\bx{\bm{x}}
\def\by{\bm{y}}
\def\bk{\bm{k}}
\thispagestyle{empty}

\begin{titlepage}

\hbox to \hsize{\YUKAWAmark \hfill YITP-96-29}
\rightline{KUNS 1381}
\rightline{September 1996}

\vspace{2cm}

\begin{center}\large\bf
Evolution of Cosmological Perturbations \\
during Reheating 
\end{center}

\bigskip

\begin{center}
Takashi Hamazaki\footnote{email address: hamazaki@murasaki.scphys.kyoto-u.ac.jp}
\end{center}

\begin{center}\it
Department of Physics, Faculty of Science, Kyoto University,\\
Kyoto 606-01, Japan
\end{center}

\begin{center}
and
\end{center}

\begin{center}
Hideo Kodama\footnote{email address: kodama@yukawa.kyoto-u.ac.jp} 
\end{center}

\begin{center}\it
Yukawa Institute for Theoretical Physics, Kyoto University, \\
Kyoto 606-01, Japan\\
\end{center}

\bigskip
\bigskip
\begin{center}\bf Abstract\end{center}
The behavior of scalar perturbations on superhorizon scales during the
reheating stage is investigated by replacing the rapidly oscillating
inflaton field by a perfect fluid obtained by spacetime averaging and
the WKB approximation. The influence of the energy transfer from the
inflaton to radiation on the evolution of the Bardeen parameter is
examined for realistic reheating processes.  It is shown that the
entropy perturbation generated by the energy transfer is negligibly
small, and therefore the Bardeen parameter is conserved in a good
accuracy during reheating. This justifies the conventional
prescription relating the amplitudes of quantum fluctuations during
inflation and those of adiabatic perturbations at horizon crossing in
the post-Friedmann stage.

\end{titlepage}

\section{Introduction}

In the inflationary paradigm cosmological large scale structures such
as galaxies and their distribution are thought to be formed from seed
density perturbations produced by quantum fluctuations of an inflaton
field\cite{r:qfluct,Bardeen.J&Steinhardt&Turner1983}.
In this scenario one can in principle determine the statistical
properties of the present large scale structure of the universe by
calculating the amplitude and the spectrum of the seed perturbations
and tracing their evolution, if the fundamental laws of nature are
specified. The former task is relatively easy and the result can be
put into a simple formula which is valid for a wide variety of the
inflaton potential. For simplified evolutionary universe models, it is
also the case for the second task as far as the linear evolutionary
stage of perturbations is concerned. This simplification is brought
about by the fact that a gauge invariant variable called the Bardeen
parameter is conserved with a good accuracy for growing modes in
Friedmann stages and the inflationary stage%
\cite{Bardeen.J&Steinhardt&Turner1983,Friemann.J&Turner1984,Brandenberger.R&Kahn1984,%
Kodama.H&Sasaki1984}.
 For example, if the reheating at the end of the inflationary stage is
instantaneous, one can determine the amplitudes of perturbations when
they reenter the horizon in the post-Friedmann stage from those of the
quantum fluctuations on the Hubble horizon scales during inflation
simply by matching the values of the Bardeen parameter. 

This powerful conservation law of the Bardeen parameter still holds
for smooth transition from the inflationary to the Friedmann stage,
provided that the equation of state of the cosmic matter changes
slowly with the cosmic expansion and that entropy perturbations can be
neglected\cite{Bardeen.J&Steinhardt&Turner1983}.  In realistic models
of reheating, however, it is not clear whether these conditions are
satisfied or not because matters with different dynamical properties
such as the inflaton field and radiation coexist possibly for a long
while. Further, according to the perturbation theory of
multi-component systems, the energy transfer processes, for example,
from the inflaton to radiation, themselves may produce additional
entropy perturbations if the energy transfer rate depends on the
energy densities or the cosmic expansion rate.  For example, it is
proposed recently that a parametric resonance of the coherent inflaton
field and other massless scalar fields may contribute as the dominant
energy transfer process in the early phase of
reheating\cite{Traschen.J&Brendenberger1990,Kofman.L&Linde&Starobinsky1994,%
Shtanov.Y&Traschen&Brandenberger1995,Yoshimura.M1995}.  In this case
large entropy perturbations can be produced by the perturbations of
the energy transfer rate since this rate for the parametric resonance
is very sensitive to the amplitude of the inflaton oscillation and the
cosmic expansion rate.

In the present paper we investigate this problem in detail and examine
whether the Bardeen parameter is well conserved or not during the
reheating phase by evaluating the entropy perturbations produced by
realistic reheating processes with the help of the gauge invariant
formalism for multi-component systems\cite{Kodama.H&Sasaki1984}. On
the basis of the result in our previous
work\cite{Kodama.H&Hamazaki1996} that the behavior of perturbations in
the stage dominated by an oscillatory inflaton $\phi$ coincides with
that of a perfect fluid obtained by a spacetime averaging and the WKB
approximation except for a sequence of negligibly short intervals
around the zero points of $\dot\phi$, we consider a system consisting
of gravity, radiation and a perfect fluid corresponding to the
inflaton.

The paper is organized as follows. First in the next section we give
the basic assumptions and explain their motivations. In particular the
behavior of a scalar field in the flat spacetime which decays to a
massless scalar particles via the standard one particle decay
process(Born decay) is briefly analyzed to prove the preservation of
coherence of the decaying inflaton field.  This analysis is also used
to determine the structure of the energy transfer term. In \S3 the
evolution equations for perturbations during reheating are given by
applying the gauge-invariant formalism for the perturbations of a
multi-component system to our system. Then in \S4 the amplitudes of
entropy perturbations produced during reheating are estimated with the
helps of these evolution equations to show that the Bardeen parameter
is conserved with a good accuracy for perturbations corresponding to
the present large scale structures. \S5 is devoted to conclusion and
discussion. The gauge-invariant perturbation theory of a
multi-component system used in \S3 is reviewed in Appendix A. The
definitions and formulae are given in a generic form in order to
correct errors in the corresponding equations in the original
article\cite{Kodama.H&Sasaki1984}.  In Appendix B a general method to
get upper bounds on solutions to a first-order differential equation
system is explained. It is used to find upper bounds on the entropy
perturbations in \S4.

Throughout this paper, the natural units $c=\hbar=1$ are adopted and
$8 \pi G$ is denoted as $\kappa^2$.  Further the notations for the
perturbation variables adopted in the
article\cite{Kodama.H&Sasaki1984} are used and their definitions are
sometimes omitted except for those newly defined in this paper.

\section{Fundamental assumptions and preliminary considerations}

In order to investigate the evolution of perturbations during the
reheating phase, we must treat a coupled system of the inflaton,
matter and gravitational field. In this paper, as the cosmic matter,
we only consider radiation whose dynamics is described by the
energy-momentum tensor of a perfect fluid
\Beqr
&&\tilde  T_{(r)\mu\nu}=(\tilde \rho_r + \tilde P_r)
\tilde u_{(r)\mu} \tilde u_{(r)\nu}+\tilde g_{\mu\nu}\tilde P_r,\\
&& \tilde P_r={1\over 3}\tilde\rho_r,
\Eeqr
where and in the followings the tilde denotes perturbed quantities.
 Further we assume that the inflaton is well described by a classical
 and coherent real scalar field $\tilde \phi$ minimally coupled with
 gravity, and its energy-momentum tensor is given by
\Beq
\tilde T_{(\phi)\mu\nu}=\partial_\mu\tilde\phi\partial_\nu\tilde\phi
-{1\over2}\tilde g_{\mu\nu}\left[(\tilde\nabla\tilde\phi)^2+U(\tilde\phi)
\right].
\Eeq
Due to the decay of the inflaton these energy-momentum tensors are
not conserved separately:
\Beqr
&& \tilde\nabla_\nu \tilde T_{ (\phi)\mu }^\nu=\tilde Q_{(\phi)\mu}
\equiv -\tilde Q_\mu,\\
&& \tilde\nabla_\nu \tilde T_{ (r)\mu }^\nu=\tilde Q_{(r)\mu}
\equiv \tilde Q_\mu.
\Eeqr

\subsection{Coherence of the inflaton}

One subtle point in the assumptions above is the requirement of
coherence on the inflaton field. If the interaction of the inflaton
field and radiation is neglected, no problem arises when we assume
that it is described by a coherent classical field. On the other hand,
when the interaction is taken into account, the consistency of this
assumption is no longer obvious because the inflaton field by itself
is a quantum field and its decay into radiation is a quantum process
in general. Further such quantum nature should be properly taken into
account when one determines the structure of the energy transfer term
$\tilde Q_\mu$.

Though we cannot justify this assumption for a generic case, it seems
reasonable at least in the case in which the potential of the scalar
field is quadratic in $\phi$ from the following observation.

Let us consider a massive scalar field $\phi$ on a flat background
interacting with a massless scalar field $\chi$ which plays the role
of radiation in the realistic situations.  For simplicity we assume
that the interaction of these fields is cubic and their Lagrangian
density is given by
\Beq
\L=-{1\over2}\left[(\partial\phi)^2+m^2\phi^2+(\partial\chi)^2
+2\mu\phi\chi^2\right].
\Eeq
Let us quantize these scalar fields by the standard canonical
quantization and introduce the creation and annihilation operators,
$(A_{\bm{p}},A_{\bm{p}}^\dagger)$ and
$(B_{\bm{k}},B_{\bm{k}}^\dagger)$, in the Schr\"{o}dinger picture by
\Beqr
&&\phi(\bm{x})=\sum_{\bm{p}}{1\over\sqrt{2V\omega_p}}
\left(A_{\bm{p}}+A_{-\bm{p}}^\dagger\right)e^{i\bm{p}\cdot\bm{x}},\\
&&\pi_\phi(\bm{x})=-i\sum_{\bm{p}}\sqrt{\omega_p\over 2V}
\left(A_{\bm{p}}-A_{-\bm{p}}^\dagger\right)e^{i\bm{p}\cdot\bm{x}},\\
&&\chi(\bm{x})=\sum_{\bm{k}}{1\over\sqrt{2V|\bm{k}|}}
\left(A_{\bm{k}}+A_{-\bm{k}}^\dagger\right)e^{i\bm{k}\cdot\bm{x}},\\
&&\pi_\chi(\bm{x})=-i\sum_{\bm{k}}\sqrt{|\bm{k}|\over 2V}
\left(A_{\bm{k}}+A_{-\bm{k}}^\dagger\right)e^{i\bm{k}\cdot\bm{x}},
\Eeqr
where $\pi_\phi$ and $\pi_\chi$ are the conjugate momentums for
$\phi$ and $\chi$, respectively, and 
\Beq
\omega_p:=\sqrt{|\bm{p}|^2+m^2}.
\Eeq
Then these canonical variables are represented on the Fock space
$\H=\H_\phi\otimes \H_\chi$ spanned by the basis vectors
\Beq
|\bm{n},\bm{m}>=\left(\sum_{\bm{n}}{1\over\sqrt{\bm{n}!}}
(A^\dagger)^{\bm{n}}\right)\Omega_\phi
\otimes\left(\sum_{\bm{m}}{1\over\sqrt{\bm{m}!}}
(B^\dagger)^{\bm{m}}\right)\Omega_\chi,
\Eeq
where $\Omega=\Omega_\phi\otimes\Omega_\chi$ is the Fock
vacuum defined by
\Beq
A_{\bm{p}}\Omega_\phi=0, \quad B_{\bm{k}}\Omega_\chi=0.
\Eeq

In this representation the free part of the Hamiltonian for each
component is written in the standard diagonal form:
\Beq
:H_\phi: =\sum_{\bm{p}}\omega_p A^\dagger_{\bm{p}}A_{\bm{p}},
\qquad
:H_\chi: =\sum_{\bm{k}}|\bm{k}| B^\dagger_{\bm{k}}B_{\bm{k}}.
\Eeq
Hence $\Omega$ is the ground state of the free part of the
Hamiltonian, but not of the total Hamiltonian. Though this implies
that $\Omega$ is unstable against the time evolution, it is a
convenient cyclic vector for constructing coherent states.

Let us consider a state $\Phi(t)$ which is represented at the initial
time $t=0$ as
\Beq
\Phi(0)=\Phi_\phi\otimes \Omega_\chi.
\Eeq
Then by the perturbation theory, taking account of the mass
renormalization, the expectation value of the annihilation operator
for $\phi$ is calculated as
\Beq
<\Phi(t)|A_{\bm{p}}|\Phi(t)>=e^{- {t\over\tau_p} }
<A_{\bm{p}}(t)>_0 + \Order{\mu^3},
\label{EV:A}\Eeq
where $A_{\bm{p}}(t)$ is the operator defined by
\Beq
A_{\bm{p}}(t):=A_{\bm{p}}e^{-i\omega_p t} 
-{i\over \omega_p\tau_p}
A_{-\bm{p}}^\dagger \cos \omega_pt,
\Eeq
$<X>_0$ denotes the expectation value of $X$ in the state $\Phi_\phi$,
and $\tau_p$ is the life of $\phi$-particles with a momentum $\bm{p}$
given by
\Beq
\tau_p={\mu^2\over 16\pi\omega_p}.
\Eeq
The expectation value of the creation operator is just given by the
complex conjugate of the above expression.

Next we consider the expectation values of products of creation and
annihilation operators.  It is convenient to define $ << X >>_t $ by
\begin{equation}
<< X >>_t  :=
 <\Phi(t)| X |\Phi(t)>
-<\Omega(t)| X |\Omega(t)>
     ,
\end{equation}
where $\Omega(t)$ is the solution to the Schr\"{o}dinger equation with
the initial condition $\Omega(0)=\Omega$. This subtraction of the
expectation with respect to $\Omega(t)$ is to eliminate the effect of
the instability of the perturbative vacuum. In this notation, the
expectation values of products of creation and annihilation operators
are given by
\Beqr
&& << A_{\bm{p}}A_{\bm{q}} >>_t
=e^{-t({1\over \tau_p}+{1\over\tau_q})}
<:A_{\bm{p}}(t)A_{\bm{q}}(t):>_0 + \Order{\mu^3},
\label{EV:AA}\\
&& << A_{\bm{p}}^\dagger A_{\bm{q}} >>_t
=e^{-t({1\over \tau_p}+{1\over\tau_q})}
<:A_{\bm{p}}^\dagger(t)A_{\bm{q}}(t):>_0 + \Order{\mu^3}.
\label{EV:AA*}\Eeqr

In particular if we take the initial state $\Phi_\phi$ as a coherent
state given by
\Beq
\Phi_\phi=\exp\left(-{1\over2}\sum_{\bm{p}}|c_p|^2\right)
\exp\left(\sum_{\bm{p}}c_p A_{\bm{p}}^\dagger\right)\Omega_\phi,
\Eeq
the expectation values of $A_{\bm{p}}$ and $A_{\bm{p}}^\dagger$, and
their products with respect to this initial state are simply given by
\Beqr
&& <A_{\bm{p}}>_0 =c_p, \quad <A_{\bm{p}}^\dagger>_0=\bar c_p,\\
&& <A_{\bm{p}}A_{\bm{q}}>_0=c_p c_q, \quad
<A_{\bm{p}}^\dagger A_{\bm{q}}>_0=\bar c_p c_q.
\Eeqr
Hence from \Eq{EV:A}, \Eq{EV:AA} and \Eq{EV:AA*} we obtain
\Beq
<< X(\bm{x})Y(\bm{y}) >>_t
\simeq <\Phi(t)|X(\bm{x})|\Phi(t)><\Phi(t)|Y(\bm{y})|\Phi(t)>,
\Eeq
up to $\Order{\mu^2}$ where $X$ and $Y$ are any of $\phi$ and $\pi_\phi$.

This result shows that the coherence of the scalar field is preserved
by its Born decay at least within a time shorter than the decay life
$\tau$. Though the above argument does not apply to the energy
transfer by the parametric resonance, it is reasonable to assume the
coherence of the inflaton in that case as well because the parametric
resonance occurs only when the inflaton field has a good coherence.

\subsection{Energy-momentum transfer term}

The simplified model analysis in the previous subsection can be used
to determine the structure of the energy-momentum transfer term
$\tilde Q_\mu$ as well.

In that model the divergence of the energy-momentum tensor of the
quantum $\phi$-field is given by
\Beq
\partial_\nu T_{(\phi) \mu}^\nu=\mu  \chi^2\partial_\mu\phi-\delta m^2
\phi\partial_\mu\phi,
\Eeq
where the second term on the right-hand side is the mass counter term.
Calculation of the expectation value with respect to $\Phi(t)$ in the
order $\mu^2$ yields
\Beqr
&Q_{(\phi) \mu}&=<< \partial_\nu T_{ (\phi)\mu  }^\nu >>_t
\nonumber\\
&&=-{\mu^2\over 16\pi}
<:\partial_\mu \hat\phi(t,\bm{x})
\sum_{\bm{p}}{1\over \omega_p\sqrt{V}}
\hat\pi_p(t)e^{i\bm{p}\cdot\bm{x}}:>_0,
\Eeqr
where 
\Beqr
&&\hat \phi(t,\bm{x}):=\sum_{\bm{p}}{1\over\sqrt{2V\omega_p}}
\left(A_{\bm{p}}e^{-i\omega_p t}+A_{-\bm{p}}^\dagger e^{i\omega_p t}
\right)e^{i\bm{p}\cdot\bm{x}},\\
&&\hat\pi_p(t):=-i\sqrt{2\omega_p}\left(A_{\bm{p}}e^{-i\omega_p t}
-A_{-\bm{p}}e^{i\omega_p t}\right).
\Eeqr
In the case in which the initial coherent state $\Phi_\phi$ contains
only particles with small momentums, as in the case of inflaton, this
expression is approximately written as
\Beqr
&Q_{(\phi) \mu} &
= -\gamma <:\partial_\mu\hat\phi(t,\bm{x})\hat\pi(t,\bm{x}):>_0
\nonumber\\
&& \simeq -\gamma \partial_\mu 
<\hat\phi(t,\bm{x})>_0  <\hat\pi(t,\bm{x})>_0
\equiv -\gamma \partial_\mu\phi \pi,
\Eeqr
where
\Beq
\gamma := {\mu^2\over 16\pi m}.
\Eeq

This result suggests that in curved spacetimes in general the energy
transfer term for the Born decay has the form
\Beq
\tilde Q_\mu =
- \tilde Q_{(\phi) \mu} = 
\partial_\mu \tilde \phi \tilde \Gamma(\tilde \phi, 
\tilde n^\nu \partial_\nu\tilde\phi),
\label{EnegyTranserTermByphi}\Eeq
where $\tilde n^\mu$ is some timelike unit vector which coincides with
the unit normal to the constant time hypersurfaces in the spatially
homogeneous case. Though we cannot determine this unit vector, its
choice has no effect in the framework of the linear perturbation
theory for the following reason. In the linear perturbation, since
$\tilde n^j$ is a first-order quantity, the perturbation of $\tilde
n^\mu \partial_\mu\tilde\phi$ depends only on $\delta n^0$ as
\Beq
\delta(\tilde n^\mu \partial_\mu\tilde\phi)=\partial_0 \delta\phi
+\delta n^0\partial_0 \phi.
\Eeq
On the other hand, for the same reason, $\delta n^0$ is determined
only by the perturbation of $\tilde g_{00}$ as
\Beq
\delta n^0=-{1\over 2}{\delta g_{00}\over g_{00}}.
\Eeq
Hence the freedom of $\tilde n^\mu$ has no effect.

The above argument on the structure of $\tilde Q_\mu$ may not apply to
the energy transfer term for other processes such as the parametric
resonance.  However, since $\partial_\mu \tilde \phi$ is the unique
vector field constructed from $\phi$ in the inflaton dominated stage,
it is reasonable to assume that ${\tilde Q}_\mu$ has the same
structure as that given in Eq.(\ref {EnegyTranserTermByphi}) for such
cases as well, though $\tilde \Gamma$ may depend on higher derivatives
of $\tilde \phi$

\subsection{WKB approximation and replacement of the scalar field by a perfect fluid}

 From the discussion so far we can formulate the evolution equation
for perturbations during the reheating stage by applying the
gauge-invariant formalism for a general multi-component system to the
system consisting of the classical scalar field, radiation and the
gravitational field. However, the equations obtained by this procedure
is rather difficult to analyze because some of the terms in the
equations becomes very large periodically when $\dot\phi$ vanishes.

In order to avoid this difficulty and make the problem tractable, we
utilize the result of our previous paper\cite{Kodama.H&Hamazaki1996}.
It is shown there that during the stage in which the rapidly
oscillating classical scalar field dominates the energy density of the
universe and its behavior is well described by the WKB form
\Beq
\tilde \phi = F(\tilde S, x),
\Eeq
where $\tilde S$ is a rapidly oscillating phase function, the behavior
of superhorizon scale perturbations coincides with that for a perfect
fluid system obtained by a spacetime averaging of the energy-momentum
tensor of $T_{(\phi)}^\mu{}_\nu$ over Hubble horizon scales except for
a sequence of negligibly short intervals around the zero points of
$\dot\phi$. This implies that we can replace the rapidly oscillating
scalar field by a perfect fluid in the investigation of the behavior
of superhorizon perturbations as far as the perturbation variables
averaged over the oscillation period are concerned. On the basis of
this result we consider a perfect fluid instead of treating the
classical scalar field directly.

The energy-momentum tensor of the perfect fluid corresponding to the
classical scalar field is given by\cite{Kodama.H&Hamazaki1996}
\Beqr
&&\tilde T_{(f)}^\mu{}_\nu=(\tilde \rho_f + \tilde P_f)\tilde u_{(f)}^\mu
\tilde u_{(f)\nu} +\tilde P \delta^\mu_\nu,\\
&& \tilde P_f=w_f \tilde \rho_f;  \quad w_f={n-2\over n+2},
\Eeqr
where we have assumed that the potential of the scalar field is given
by a simple power-law function
\Beq
U={\lambda\over n}|\phi|^n.
\Eeq
We assume this power law form throughout this paper.  $\tilde \rho_f$
and $\tilde u_{(f)}^\mu$ is represented in terms of the original
scalar field as
\Beqr
&&\tilde \rho_f={n+2\over 2n}<(\nabla\tilde\phi)^2>
\simeq {n+2\over 2n}<(\partial_{\tilde S} F)^2>(\tilde \nabla<\tilde S>)^2,\\
&& \tilde u_{(f)\mu}=
{\partial_\mu<\tilde S>\over [ (\tilde\nabla<\tilde S>)^2 ]^{1 \over 2} },
\Eeqr
where $<X>$ represents a spacetime average of $X$ over the Hubble
horizon scales. This approximation is good if the parameter defined by
\Beq
\epsilon^2:={<(\nabla F)^2>\over <|\partial_S F|^2(\nabla S)^2>}
\Eeq
is much smaller than unity. Since $\epsilon$ represents the ratio of
the cosmic expansion rate $H$ to the oscillation frequency of the
scalar field, this condition is satisfied in the rapidly oscillating
phase.

In order to formulate the perturbation equations for this perfect
fluid and radiation, we must rewrite the spacetime average of the
energy-momentum transfer term (\ref{EnegyTranserTermByphi}) in terms
of the fluid variables. If we write $<\tilde Q_\mu>$ as $\tilde Q_\mu$
for simplicity, it must have the form
\Beq
\tilde Q_\mu=\partial_\mu <\tilde S> R(<\tilde\phi>^2,
\tilde n^\nu \partial_\nu <\tilde S>).
\Eeq
 From the argument in the previous subsection we can take
$\partial^\mu <\tilde S>$ as $\tilde n^\mu$. Hence, from the
relativistic virial theorem\cite{Kodama.H&Hamazaki1996}
\Beq
<U(\tilde\phi)>=-{1\over n}<(\tilde \nabla\tilde\phi)^2>
\left[1+\Order{\epsilon}\right],
\Eeq
it is simply written as
\Beq
\tilde Q_\mu=\tilde u_{(f)\mu}{\tilde Q}
      ,
\hspace{0.5cm}
{\tilde Q}=
G (\tilde\rho_f)
      .
\Eeq
 For example, for the Born decay in the model considered in the
previous subsection, $G$ is simply given by
\Beq
G (\tilde\rho_f)
={2n\over n+2}\gamma \tilde \rho_f.
\Eeq

On the other hand, for the energy transfer by the parametric
resonance, $\tilde Q$ should be modified as
\begin{equation}
 {\tilde Q}=
G({\tilde \rho}_f, {1 \over 3}{\tilde \nabla}_\mu {\tilde u}^\mu_f)
  ,
\end{equation}
where ${\tilde \nabla}_\mu {\tilde u}^\mu_f /3$ represents the
expansion rate of the $\phi={\rm const}$ hypersurface, and coincides
with the Hubble parameter $H$ in the unperturbed background.  This
dependence arises because the duration of the parametric resonance for
each mode of massless fields coupled with $\phi$ depends on the cosmic
expansion rate.  Though ${\tilde \nabla}_\mu {\tilde u}^\mu_f$ is of
order $\epsilon$ in the WKB approximation scheme, it may not be
neglected because ${\tilde Q}$ has a strong dependence on it for the
parametric resonance decay.

\section{Evolution equations for perturbations}

Under the assumptions given in the previous section, we can easily
write down the evolution equation of perturbations during reheating by
applying the gauge invariant perturbation theory of a multicomponent
system to the current system.  Basically a perturbation of this system
is described by the gauge invariant density contrasts and the shears
of the 4-velocity for the inflaton fluid and radiation.  But in order
to investigate the dynamical behavior of the Bardeen parameter, it is
more convenient to choose variables which respect the decomposition of
perturbations into the adiabatic modes and the entropy modes.  Hence,
as the basic variables, we adopt the curvature perturbation $\Phi$,
the total shear velocity $V$, and the quantities representing the
difference of the density contrasts and the shear velocities between
the inflaton and radiation, $Y$ and $Z$, defined by
\begin{equation}
 Y = {\rho_f \rho_r \over \rho^2}S_{f r},
\hspace{0.5cm}
 Z = {\rho_f \rho_r \over \rho^2}{aH \over k} V_{f r}.
\label {e:defent}
\end{equation}
Note that $Y$ and $Z$ become zero at the beginning and at the end of
the reheating stage when the energy density of radiation or the
inflaton is negligible.

With the help of the general formulae given in Appendix A, we can
easily write down the evolution equations for the basic variables by
specializing the space dimension to 3.  First, from Eq.(\ref {e:Ds}),
(\ref {e:Vs}), (\ref {e:gamcom}), and (\ref {e:gamrel}), the evolution
equations for $\Phi$ and $\Upsilon$ are written in terms of these
variables as
\begin{eqnarray}
&&
{\cal D} \Phi + \Phi = -{3 \over 2} I_K (1+w) \Upsilon ,
\label {e:evophi}   \\
&&
{\cal D} \Upsilon + {3 \over 2} I_K (1+w) \Upsilon 
      -{K \over a^2 H^2} \Upsilon 
\nonumber \\
&& \quad \quad
=
  \left[
 -1+{2 \over 3} \left(k \over a H\right)^2 {C_K \over I_K}
       {1 \over 1+w} {1 \over h} \left(w_f h_f +{1 \over 3}h_r\right)
  \right]
  \Phi
\nonumber \\
&& \quad \quad \quad
+ {4 \over 3}{1+w_f \over (1+w)^2}\left(w_f-{1 \over 3}\right)Y
    ,
\label {e:evoups}
\end{eqnarray}
where ${\cal D}= a ( d / da )=d / d({\log a})$.  On the other hand, by
taking account of Eqs. (\ref {e:Fcalp}), (\ref {e:EMconp}), and (\ref
{e:momdif}), Eq.(\ref {e:ED}) and (\ref {e:EV}) reduce to the
following evolution equations for $Y$, $Z$:
\begin{eqnarray}
&&{\cal D}Y + 3\left({1 \over 3}+w_f-2w\right)Y =
 - \left({k \over aH}\right)^2 Z 
 -{3 \over 4}{Q \over H \rho}{1+w \over 1+w_f}E_{c f}
\nonumber \\
&& \quad \quad
 +{2Q \over 3H \rho}{1 \over 1+w_f}\left( 1+{3 \over 4}w_f \right)
   {C_K \over I_K}\left({k \over a H}\right)^2 \Phi,
\label {e:evoY} \\
&& 
{\cal D}Z 
            +\left[
   {3 \over 2}(1+w)I_K-{K \over a^2 H^2}
        +3(w_f-2w)+(1-3 w_f){h_r \over h}\right.
\nonumber \\  
&& \quad \quad
        \left.  +{Q \over Hh}\left(
          w_f + { 2 \over 3 }
          +{4 \over 3}{1 \over 1+w_f}{\rho_r \over \rho_f}
               \right)
             \right]Z
\nonumber \\   
&& \quad \quad =
   {1 \over h}\left[
           {4 \over 3} w_f \rho_r
           +{1 \over 3}(1+w_f) \rho_f
             \right] Y
\nonumber \\
&& \quad \quad
     +{2 \over 3}\left(w_f -{1 \over 3}\right){C_K \over I_K}
     \left({k \over a H }\right)^2 
     {\rho_f \rho_r \over \rho^2}{\Phi \over 1+w}
    .
\label {e:evoZ}
\end{eqnarray}
In particular, subtracting Eq.(\ref {e:evoups}) from Eq.(\ref
{e:evophi}), we obtain
\begin{eqnarray}
&& 
 {\cal D} (\Phi - \Upsilon) =
  - {K \over a^2 H^2} \Upsilon
  - {2 \over 3} \left({k \over a H}\right)^2 {C_K \over I_K}
     {1 \over 1+w} {1 \over h} \left(w_f h_f  +{1 \over 3}h_r \right)
       \Phi
\nonumber \\
&& \quad \quad
   - {4 \over 3}{1+w_f \over (1+w)^2}\left(w_f-{1 \over 3}\right)Y,
\label {e:Bard}
\end{eqnarray}
where $\Phi - \Upsilon$ denoted as $\zeta$ in the article\cite{r:Mukh}
is referred to as the Bardeen parameter from now
on\cite{Bardeen.J&Steinhardt&Turner1983}.  Taking into account that
$\Phi - \Upsilon$ and $\Phi$ are of the same order, and that the
spatial curvature $K$ is practically zero because of the inflationary
expansion, we immediately confirm from this equation that the Bardeen
parameter $\Phi - \Upsilon$ is conserved on superhorizon scales, if
the entropy perturbation $Y$ is negligibly small.  In the next
section, we will investigate in how good accuracy the Bardeen
parameter is conserved by evaluating the amplitude of the entropy
perturbation generated in realistic reheating processes.

In the evolution equations of the entropy perturbation, the
perturbation of the energy transfer rate works as the source
term. When the energy transfer rate ${\tilde Q}$ depends only on
${\tilde \rho}_f$, the perturbation of the energy transfer rate is
expressed in terms of the basic variables as
\begin{eqnarray}
E_{c f} &=&
 { G_{\rho_f} (\rho_f) \rho_f \over G(\rho_f) }
\Delta_{c f}
\nonumber \\
           &=&
 { G_{\rho_f} (\rho_f) \rho_f \over G(\rho_f) }
  {1 + w_f \over 1+w}
    \left[  {2 \over 3}{C_K \over I_K}
        \left({k \over aH}\right)^2 \Phi
        + {4 \over 3}{\rho \over \rho_f}Y 
    \right],
\label {e:enerho}
\end{eqnarray}
where $G_{\rho_f}$ denotes the partial derivative of $G$ with respect
to ${\rho_f}$.  If the energy transfer rate depends also on the cosmic
expansion rate ${\tilde \nabla}_\mu {\tilde u}^\mu_f$ as
\begin{equation}
 {\tilde Q}=
G({\tilde \rho}_f, {1 \over 3}{\tilde \nabla}_\mu {\tilde u}^\mu_f),
\end{equation}
the following term should be added to the right hand side of Eq.({\ref
{e:enerho}}):
\begin{eqnarray}
&&
{G_H (\rho_f, H) ~ H \over G (\rho_f, H) }
\left( {1 \over 3}{k \over a H }V_f - {K \over a^2 H^2}\Upsilon \right)
\nonumber \\
&& \quad \quad =
{G_H (\rho_f, H) ~ H \over G (\rho_f, H) }
\left[
  {1 \over 3}\left({k \over a H }\right)^2 \Upsilon
    + {4 \over 9} {1 \over 1+w}\left({k \over a H }\right)^2 
    {\rho \over \rho_f }Z - {K \over a^2 H^2}\Upsilon 
\right].
\label {e:traHub1}
\end{eqnarray}
On the other hand, if ${\tilde \nabla}_\mu {\tilde u}^\mu$, or
${\tilde \nabla}_\mu {\tilde u}^\mu_r$ are adopted as the local Hubble
constant, the terms to be added are given by
\begin{eqnarray}
&&
{G_H (\rho_f, H)  ~ H \over G (\rho_f, H) }
\left(
  {1 \over 3}{k \over a H }V  - {K \over a^2 H^2}\Upsilon
\right)
\nonumber \\
&& \quad \quad =
{G_H (\rho_f, H) ~ H \over G (\rho_f, H) }
\left[
  {1 \over 3}\left({k \over a H }\right)^2 \Upsilon
    - {K \over a^2 H^2}\Upsilon
\right],
\end{eqnarray}
and
\begin{eqnarray}
&&
{G_H  (\rho_f, H) ~ H \over G (\rho_f, H) }
\left(
  {1 \over 3}{k \over a H }V_r - {K \over a^2 H^2}\Upsilon
\right)
\nonumber \\
&&  \quad \quad =
{G_H (\rho_f, H) ~ H \over G (\rho_f, H) }
\left[
  {1 \over 3}\left({k \over a H }\right)^2 \Upsilon
    - {1 \over 3} {1+w_f \over 1+w} \left({k \over a H }\right)^2
    {\rho \over \rho_r }Z - {K \over a^2 H^2}\Upsilon
\right],
\label {e:traHub2}
\end{eqnarray}
respectively.

In all of these expressions for $E_{c f}$, all the terms in proportion
to $\Phi$ or $\Upsilon$ are multiplied by coefficients of order $ ({k
/ a H})^2 $.  In the next section, we will show that this suppression
factor makes the contribution of the entropy perturbations negligibly
small, even if we take into account a possible dynamical growth of
$Y$ and $Z$.

\section{Conservation of the Bardeen parameter during the reheating stage}

 From Eq.(\ref {e:Bard}), it follows that the entropy perturbation 
does not affect the conservation of the Bardeen parameter 
if the condition
\begin{equation}
\int_{a_s}^{a_e} { da \over a }
 |Y| 
\ll
|\Phi|
\label {e:purpo}
\end{equation}
is satisfied, where $a_s$ and $a_e$ are the values of the scale factor
at the start and at the end of the reheating stage.
In this section we show that this condition are satisfied
in the realistic chaotic inflation scenario whose 
dominant reheating processes are the parametric resonance 
and/or the Born decay.

 First we define the index $g_{\rho_f}$ and $g_H$ by 
\begin{equation}
g_{\rho_f} :=
  {G_{\rho_f} {\rho_f} \over G}
      ,
\hspace{0.5cm}
g_H :=
  {G_H  H \over G}
      .
\end{equation}
Though $g_{\rho_f}$ is bounded by unit from below,
\begin{equation}
g_{\rho_f} \ge 1,
\end{equation}
for realistic reheating processes, it may become much larger than unity
in the stage in which the parametric resonance is effective.  It is difficult 
to evaluate the upper bound on $g_{\rho_f}$ in this stage because we have poor 
knowledge on the strong parametric resonance.
However, since the analysis in the weak parametric resonance\cite{Shtanov.Y&Traschen&Brandenberger1995} 
shows that $G$ behaves as $G \sim \exp ( {\rm O}(g_{\rho_f}) )$ when 
$g_{\rho_f} \gg 1$, it is expected that $g_{\rho_f}$ does not exceed unity
by many orders of magnitude. For example, according to the recent 
numerical investigation taking account of rescattering of produced 
particles\cite {r:param4}, the effective value of $g_{\rho_f}$ does not 
exceed $100$. In contrast to $g_{\rho_f}$, $g_H$ does not have a definite sign 
and may becomes negative. However, from the analysis of weak parametric 
resonance, it is expected that its absolute value $|g_H|$ is at most 
of the same order as $g_{\rho_f}$. Hence, when the distinction of these is not 
important, we use  
\begin{equation}
g := \max \{ g_{\rho_f}, |g_H|  \}
      .
\end{equation}
Note that as the Born decay dominates in the energy transfer, $g_H$ vanishes
and $g_{\rho_f}=g$ approaches unity.

In order to evaluate the amplitude of entropy perturbations, it is 
convenient to decompose the reheating stage into the following four
substages:
\begin{eqnarray}
&&
1)~ a_s \le a < a_1  \quad {G \over H \rho_f} \gg 1 
     ,
\nonumber \\
&&
2)~ a_1 \le a < a_2  \quad {G \over H \rho_f} g \gg 1 
         \gsim {G \over H \rho_f}
     ,
\nonumber \\
&&
3)~ a_2 \le a < a_3  \quad 1 \gsim {G \over H \rho_f} g
     , \quad 1\gsim {G\over H\rho_f},
\nonumber \\
&&
4)~ a_3 \le a < a_e  \quad {G \over H \rho_f} \gg 1 
     .
\end{eqnarray}
The first substage corresponds to the period during which an explosive energy 
transfer occurs by the parametric resonance. As the amplitude of the 
inflaton oscillation decreases and the parametric resonance gets
less effective, $G/H\rho_f$ becomes less than unity. If $g$ is much greater
than unity in this phase, the second substage appears. On the other hand,
if the Born decay already dominates at that time and $g=1$, this stage 
does not appear and the system goes directly to the third substage, during
which the Born decay is the main process of reheating but it is still slower
than the cosmic expansion. Finally as the cosmic expansion rate decreases
with time, $G/H\rho_f$ becomes greater than unity again and the reheating
completes. This is the last substage.

We evaluate the upper bound of the amplitude of the entropy
perturbation in each stage by utilizing the technique explained in
Appendix B.  We assume $K=0$ henceforth.  First we put Eqs. (\ref
{e:evoY}) and (\ref {e:evoZ}) into the matrix form
\begin{equation}
{\cal D}\bm{X}=  \bg{\Omega}\bm{X} +  \bm{S} 
   ,
\label {e:2times2}
\end{equation}
where 2-column vectors $\bm{X}$ and $\bm{S}$, and $2 \times 2$
matrix $\bg{\Omega}$ are defined by
\begin{eqnarray}
&&\bm{X} :=
   \left (  \begin{array}{c} 
                       Y \\
                       Z  \end{array}
   \right )
     ,
\nonumber \\
&&  \bg{\Omega} :=
   \left ( \begin{array}{cc}
%
                    \begin{array}{l}
                             -1-3 w_f +6w   \\
                             -{G \over H \rho_f}{g_{\rho_f}}    
                    \end{array}
&
             ({k \over a H})^2 (
                   -1-{G \over 3 H \rho_f} g_H {1 \over 1+w_f}
                               )
                         \\
                     \quad & \quad \\
     {4 \over 3}w_f{\rho_r \over h}+{1 \over 3}(1+w_f){\rho_f \over h}
& 
            \begin{array}{c}
     -{3 \over 2}+{9 \over 2}w-3 w_f-(1-3 w_f){h_r \over h} \\
          -{G \over H \rho_f}\{
       (w_f+ {2 \over 3}) {\rho_f \over h}
     +{4 \over 3}{1 \over 1+w_f}{\rho_r \over h}
                             \}
            \end{array}
        \end{array} 
    \right ) 
     ,
\nonumber \\
&&  \bm{S} :=
  \left ( \begin{array}{c}
      -{1 \over 2} {G \over H \rho} g_{\rho_f} 
        + {1 \over 6}{G \over H \rho} 
         { 4+3 w_f \over 1+ w_f } \\
\quad \\
      -{2 \over 9}{ 1 - 3 w_f \over 1+w }
      {\rho_f \rho_r \over \rho^2}
       \end{array} 
    \right )
 \left({k \over a H}\right)^2 \Phi 
\nonumber \\
&& \quad \quad \quad \quad
+
  \left ( \begin{array}{c}
       -{1 \over 4}{G \over H \rho} g_H 
         { 1+w \over 1+w_f } \\
\quad \\
       0
       \end{array} 
    \right )
 \left({k \over a H}\right)^2 \Upsilon
      .
\nonumber \\
&& \quad
\end{eqnarray}
In order to estimate the upper bound of $\bg{\Omega}_H := \bg{\Omega}
+\bg{\Omega}^\dagger$, we decompose it into a sum of three
matrices as
\begin{eqnarray}
&& \bg{\Omega}_{H1} =
  \left(
     \begin{array} {cc}
       -2-6 w_f +12w  &  0 \\  
       0              &  -3+9w-6 w_f -(2-6 w_f){h_r \over h}
     \end{array}
   \right)
   ,
\nonumber \\
&& \bg{\Omega}_{H2} =
  \left(   
     \begin{array} {cc}
       0 & 
           \begin{array}{c}
                    -({k \over a H})^2 (
                    1+{G \over 3 H \rho_f} g_H {1 \over 1+w_f}
                     ) \\
                       +
                     {4 \over 3}w_f{\rho_r \over h}
                    +{1 \over 3}(1+w_f){\rho_f \over h}
           \end{array}
                      \\
\quad & \quad \\
           \begin{array}{c}
                    -({k \over a H})^2 (
                    1+{G \over 3 H \rho_f} g_H {1 \over 1+w_f}
                     ) \\
                       +
                     {4 \over 3}w_f{\rho_r \over h}
                    +{1 \over 3}(1+w_f){\rho_f \over h}
           \end{array}
         & 0
     \end{array}
  \right)
   ,
\nonumber \\
&& \bg{\Omega}_{H3} =
  \left(
    \begin{array} {cc}
 -2{G \over H \rho_f}{g_{\rho_f}}
         & 0 \\       
       0 & 
     -{G \over H \rho_f}\{
       (2 w_f+ {4 \over 3}) {\rho_f \over h}
     +{8 \over 3}{1 \over 1+w_f}{\rho_r \over h}
                             \}
     \end{array}
  \right)
   .
\end{eqnarray}
Then the maximum eigenvalue of these matrices are given by
\begin{eqnarray}
&&  \lambda_{m1} = \max \left\{
           -2-6 w_f +12w ,
     -3+9w-6 w_f -(2-6 w_f){h_r \over h}
                        \right\}
          , 
\\
&&  \lambda_{m2} =
  -\left({k \over a H}\right)^2 
   \left(  1+{G \over 3 H \rho_f} g_H {1 \over 1+w_f} \right)  
  + {4 \over 3}w_f{\rho_r \over h}
  +{1 \over 3}(1+w_f){\rho_f \over h}
          ,
\label {e:2nd} \\
&&  \lambda_{m3} =
  - \min \left\{
       2{G \over H \rho_f}{g_{\rho_f}}  ,
    {G \over H \rho_f}\left[
       \left(2 w_f+ {4 \over 3}\right) {\rho_f \over h}
     +{8 \over 3}{1 \over 1+w_f}{\rho_r \over h}\right]
   \right\},
\end{eqnarray}
respectively.  Hence the maximum eigenvalue of $\bg{\Omega}_H$ is
bounded by the sum of them,
\begin{eqnarray}
 \lambda_m &:=&
 \lambda_{m1} +  \lambda_{m2} + \lambda_{m3}
\nonumber \\
&\le&  N
  -\left({k \over a H}\right)^2 
   \left( 1+{G \over 3 H \rho_f} g_H {1 \over 1+w_f}\right) 
\nonumber \\
&& \quad \quad
  - \min \left\{
       2{G \over H \rho_f}{g_{\rho_f}}  ,
    {G \over H \rho_f}\left[
       \left(2 w_f+ {4 \over 3}\right) {\rho_f \over h}
     +{8 \over 3}{1 \over 1+w_f}{\rho_r \over h}
    \right]
  \right\},
\label {e:lambup}
\end{eqnarray}
Here $N$ is the maximum value of the sum of $\lambda_{m1}$ and the two
last terms on the right hand side of (\ref {e:2nd}).  Since this sum
becomes maximum at $\rho_r=0$ or $\rho_f=0$ for fixed $w_f$, $N$ is
given by
\begin{equation}
N :=
 \max \{ 2-5 w_f, 6 w_f - {5 \over 3} \} 
     .
\end{equation}
The source term is evaluated as
\begin{equation}
\| \bm{S} \| \sim 
   \Order{ 
	\left(g { G \over H \rho } +1\right) 
	\left( { k \over a H } \right)^2 |\Phi| 
   },
\end{equation}
taking account of the fact that $\Phi$ is of the same order as
$\Upsilon$.
 
First in the first stage, since $\lambda_m$ is negative and its
absolute value$\sim G/H\rho_f$ is much larger than unity, Eqs.(\ref
{e:uplim}) and (\ref {e:uplimp}) in the Appendix B yields
\begin{equation}
\| \bm{X}(a) \| \le \Order{g \left( { k \over a H } \right)^2 |\Phi|}.
\end{equation}
Next in the second stage, $\lambda_m$ is bounded as $\lambda_m\le N$
since $(k/aH)^2g$ is practically much smaller than unity. Hence, by
using Eq.(\ref {e:uplim}) in the Appendix B and by taking account of
the order of magnitude of $\|\bm{X}\|$ at the end of the previous
stage, $\| \bm{X}(a_1) \|$, we obtain
\begin{equation}
\| \bm{X}(a) \| \le 
 \left( { a \over a_1 } \right)^{N / 2}  
  \Order{
  	g_1 \left( { k \over a H } \right)^2 |\Phi|
   },
\end{equation}
where $g_1$ is the value of $g$ at $a=a_1$. In the third stage, in the
same way we obtain
\begin{equation}
\| \bm{X}(a) \| \le 
 \left( { a \over a_1 } \right)^{N / 2}  
  \Order{
  g_1 \left( { k \over a H } \right)^2 |\Phi|
  }.
\end{equation}
Finally in the fourth stage, since the eigenvalue of $\bg{\Omega}$ is
negative, and its absolute value increases exponentially in time, the
influence of the previous stage on $\bm{X}(a)$, corresponding to the
first-term in Eq.(\ref{e:uplim}), is rapidly erased, and
$\|\bm{X}(a)\|$ settles down to a value determined locally by
$\|\bm{S}\|$ as 
\begin{equation}
\| \bm{X}(a) \| \le 
  \Order{ \left( { k \over a H } \right)^2 |\Phi|},
\end{equation}
as seen from Eqs.(\ref {e:uplim}) and (\ref {e:uplimp}).  This
estimate is actually a quite weak one. The actual value of
$\|\bm{X}\|$ decreases exponentially in time in this stage due to the
suppression factor $\rho_f/\rho$ in the definition
(\ref{e:defent}). Hence $Y$ and $Z$ take nonnegligible values only in
stages when the inflaton and radiation coexists, and vanish as soon as
the energy transfer completes effectively.

These estimates on the upper bound of $\| \bm{X} \|$ can be used to
obtain stronger bounds on $Y$ and $Z$. To see this, let us write the
second row of Eq.(\ref{e:2times2}) as
\begin{eqnarray}
&& {\cal D} Z = \Omega_Z Z + S_Z, \\
&& \Omega_Z := -{3 \over 2}+{9 \over 2}w-3 w_f - (1-3 w_f){h_r \over h}
          - \Order{ {G \over H \rho} },
\\
&& S_Z := \Order{Y} + \Order{\left({ k \over a H }\right)^2 \Phi },
\end{eqnarray}
and regard this as the evolution equation for $Z$. Then since $\Omega_Z$ is
bounded as
\Beqr
&&\Omega_Z \le  N_Z - \Order{{G \over H \rho} },\\
&& N_Z := \max \left\{ -3 w_f, {3 \over 2}(w_f-1) \right\},
\Eeqr
using the values of the upper bound obtained in the previous analysis
 on $|Y| ( \le \| \bm{X} \| )$ and applying Eqs.(\ref {e:uplim}) and
 (\ref {e:uplimp}) to the above equation on $Z$, we obtain the
 following stronger upper bound on $|Y|$ in the first and the fourth
 stage:
\begin{equation}
|Z| \le \Order{{ H \rho \over G } g \left({ k \over a H }\right)^2 |\Phi|}.
\end{equation}

We can apply the same method to the evolution equation for $Y$ obtained
from the first row of \Eq{e:2times2},
\begin{eqnarray}
&& {\cal D} Y = \Omega_Y Y + S_Y, \\
&& \Omega_Y := -1-3 w_f +6 w -{G \over H \rho_f}{ g_{\rho_f} },\\
&& S_Y := \Order{\left(1+ {G \over H \rho_f} g \right) 
    \left({ k \over a H }\right)^2 Z }
     + \Order{ {G \over H \rho} g \left({ k \over a H }\right)^2 \Phi}.
\end{eqnarray}
Now $\Omega_Y$ is bounded as
\Beqr
&& \Omega_Y \le N_Y -{G \over H \rho_f}{ g_{\rho_f} },\\
&& N_Y := |1-3 w_f|.
\Eeqr
Hence we obtain in the first stage
\begin{equation}
|Y(a)| \le \Order{\left({ k \over a H }\right)^2 \Phi },
\end{equation}
assuming that
\begin{equation}
 g  \left({ k \over a H }\right)^2 < 1,
\end{equation}
in the second stage
\begin{equation}
|Y(a)| \le 
\Order{
  \max \left\{ 
 1, g_1 \left({ a \over a_1 }\right)^{N/2} \left({ k \over a H }\right)^2 
       \right\} }
 \left({ k \over a H }\right)^2 \Phi,
\end{equation}
and in the third stage
\begin{equation}
|Y| \le 
 \left({ a \over a_2 }\right)^{N_Y}
\Order{
  \max \left\{ 
 1, g_1 \left({ a \over a_1 }\right)^{N/2} \left({ k \over a H }\right)^2 
       \right\} }
 \left({ k \over a H }\right)^2 |\Phi| .
\end{equation}

Putting these estimates together, we finally obtain
\begin{eqnarray}
\int_{a_s}^{a_e} { da \over a } |Y| 
&\cong&  
\int_{a_1}^{a_3} { da \over a } |Y| 
\nonumber \\
&\le& 
 \left({ a_3 \over a_1 }\right)^{N_Y}
\Order{
  \max \left\{ 
 1, g_1 \left({ a_3 \over a_1 }\right)^{N/2} \left({ k \over a H }\right)^2_3 
       \right\}
 }
 \left({ k \over a H }\right)^2_3 |\Phi|.
\end{eqnarray}       
Taking account of $N/2 \ge N_Y$ when $n \ge 2$, we can conclude that
the entropy perturbation $Y$ does not affect the conservation of the
Bardeen parameter if
\begin{equation}
 g_1
\left( { a_e \over a_s } \right)^{N_Y/2+N/4}
\left( { k \over a H } \right)^2_e \ll 1 .
\label {e:cnsrvcnd}
\end{equation}
is satisfied.  Since $g_1$ is a monotonic function of $\rho_f$ and it
is expected that it does not exceed unity much, we drop it henceforth.

The condition (\ref {e:cnsrvcnd}) gives a lower bound on the energy
density $\rho(a_e)$ at the end of reheating.  Let us evaluate its
order of magnitude for realistic values of the physical
parameters. First we introduce the parameter $y$ defined by
$\rho(a_e)^{ 1 /4 } = 10^{y} $GeV. This parameter is related to the
value of $k/aH$ at the end of reheating for perturbations
corresponding to the present cosmological structures of $10$ Mpc
scales by
\begin{equation}
\left( { k \over a H } \right)^2_e \sim 10^{-2y-15}
           .
\end{equation}
Here we have assumed $a(t_{eq}) / a(t_0) \sim 10^{-4}$ where $t_{eq}$
is the equality time.  In order to generate density perturbations
consistent with the observed anisotropy of the cosmic microwave
background ${ \delta T / T } \sim 10^{-5}$, the energy density at the
time when the relevant perturbations cross the Hubble horizon in the
inflationary stage should be given by
\begin{equation}
{\rho}^{ 1 /4 } \sim 10^{16} \r{GeV},
\end{equation}
which is approximately equal to $\rho(a_s)$. Hence from the inequality
\begin{eqnarray}
{ a_e \over a_s } \le 
\left( { \rho(a_s) \over \rho (a_e) } \right)^{ 1 / [  3(1+w_{min}) ] }
\nonumber \\
w_{min}:=\min \left\{ w_f, {1 \over 3} \right\},
\end{eqnarray}
we can see that
the condition (\ref {e:cnsrvcnd}) holds, if 
\Beq
y> \left\{\begin{array}{ll}
 \displaystyle {576w_f-356 \over 36w_f+13} & \r{for}\quad w_f\ge{1\over3},\\
 \displaystyle {221w_f-19\over 5w_f-10} & \r{for}\quad w_f\le{1\over3}
\end{array}\right.  
\label {e:solid}
\Eeq
is satisfied.

This condition on $y$ is a strict one from a mathematical point of
view.  However, it seems to be too strong from a physical point of
view because $Y$ takes nonnegligible value only around $x:=\rho_r /
\rho_f \sim 1 $, while the above estimate is based on the upper bounds
on $\lambda_m$ and $\Omega_Y$ that correspond to the values at $x=0$
or $x=\infty$.  Hence a practical condition is obtained by replacing
$N$ and $N_Y$ in Eq.(\ref {e:cnsrvcnd}) by the values of $\lambda_m$
and $\Omega_Y$ at $x=1$.  It is expressed as
\Beq
y> \left\{\begin{array}{ll}
 \displaystyle -{100w_f+404\over29w_f+57} & \r{for}\quad w_f\ge{1\over3},\\
 \displaystyle -{135w_f^2+370w_f+299\over18w_f^2+65w_f+43} & \r{for}\quad w_f\le{1\over3}.
\end{array}\right.
\label {e:dotted}
\Eeq

These conditions are depicted in Figure 1. 
\begin{figure}[htb]
   \centerline{\epsfysize=8.5cm \epsfbox{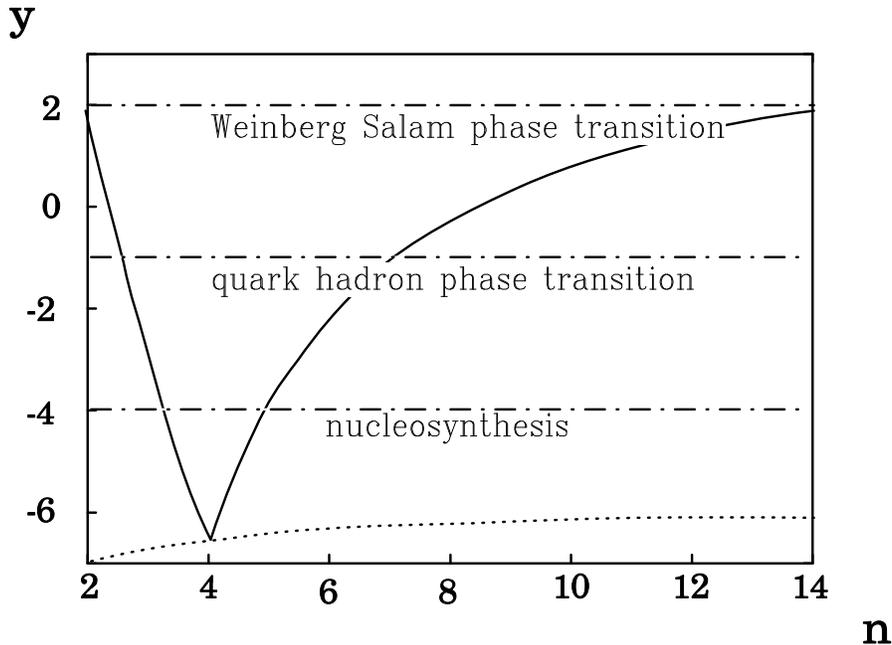}}
  \vspace {5mm}
   \caption{
   The energy density at the end of reheating 
   ($\rho^{1/4}=10^y {\rm GeV}$) vs power of the inflaton potential
   $n$
   }
   \protect\label{f:fig1}
\end{figure}
\noindent
In this figure the solid line and the dotted line correspond to (\ref
{e:solid}) and (\ref {e:dotted}), respectively, and the Bardeen
parameter is well conserved in the region above these lines. This
figure shows that reheating does not affect the conservation of the
Bardeen parameter if it terminates before the primordial
nucleosynthesis.  Hence we can conclude that in all the realistic
models based on chaotic inflation, the Bardeen parameter stays constant
in a good accuracy during reheating.  This justifies the conventional
prescription relating the amplitude of adiabatic perturbations at
horizon crossing in the post-Friedman stage to the value of the
Bardeen parameter in the inflationary stage.

\section{Discussion}

In this paper we have investigated the evolution of perturbations
during reheating taking account of the effect of the energy transfer
from the inflaton to radiation, by replacing the inflaton field by a
perfect fluid obtained by the spacetime averaging and the WKB
approximation.  By evaluating the amplitudes of entropy perturbations
generated during reheating and their influence on the adiabatic
component of perturbations, we have shown that the Bardeen parameter
is well conserved during the reheating stage as well as in the
inflationary stage and the post-Friedmann stage for realistic models. 
Though we have considered only the parametric resonance and the Born
decay as the dominant reheating processes for definiteness, the
conclusion holds rather generally since the arguments are insensitive
to the details of the models.

Of course our analysis does not exhaust all the possible models of
inflation.  In particular we have only considered the case in which
the inflaton is described by a single component field. Though a simple
multi-component extension does not seem to change the conclusion as
far as the fluid replacement of the inflaton fields during the
reheating stage gives a good approximation, some subtlety may occur if
scalar fields with very tiny masses coexist with the inflaton
fields. For example, if there exists a scalar field which affects
physical parameters controlling the reheating processes such as
particle masses or coupling constants of physical particles but has no
dynamical effect by itself during reheating, it may produce large
entropy perturbations and affect the Bardeen parameter. In particular,
as our analysis suggests, this possibility may become important if
such a field affects the parameters controlling the parametric
resonance processes. In such multi-component systems, entropy
perturbations produced during reheating may survive after reheating
and have important effects on the present universe, as discussed by
Yokoyama et al. as a mechanism to generate the baryon isocurvature
perturbation\cite{r:Yokoya}.  These problems in the multi-component
extension are under investigation.

\section*{Acknowledgments}

T. H. would like to thank Prof. H. Sato for continuous encouragements.
He would like to thank Y. Nambu, M. Sasaki, E. Stewart, and J. Yokoyama
for fruitful discussions.  This work was partly supported by the
Grant-in-Aid for Scientific Research of the Ministry of Education,
Science, Sports and Culture of Japan(H.K.:40161947).

\appendix

\bigskip
\bigskip

\noindent{\Large\bf
Appendices
}

\def\theequation{\Alph{section}.\arabic{equation}}
\def\prefixsection{}

\section{Gauge invariant Perturbation Theory of Multi Component Systems}

In this appendix, we recapitulate the basic definitions
and equations of the gauge invariant perturbation theory
for multicomponent systems developed in the article\cite {Kodama.H&Sasaki1984}.
The purpose is twofold.
The one is to explain the notations and to provide the basic equations
used in the text.
The other is to correct errors of the corresponding equations 
in \cite {Kodama.H&Sasaki1984}.

The Einstein equations and the equations of motion of a perturbed
multi-component system are given by
\Beqr
&&\tilde G^\mu_\nu=\kappa^2 \tilde T^\mu_\nu,\label{EinsteinEq:general}\\
&&\tilde\nabla_\nu \tilde T^\nu_{(\alpha)\mu}=\tilde Q_{(\alpha)\mu},
\label{EOM:general}\Eeqr
where $\tilde Q_{(\alpha)\mu}$ represents the energy-momentum transfer 
term for a component $(\alpha)$, which satisfies
\Beq
\sum_{(\alpha)}\tilde Q_{(\alpha)\mu}=0.
\label{EMTransferRate:constraint}\Eeq

In a spatially homogeneous background with metric
\Beq
ds^2=-dt^2 + a^2d\sigma_n^2,
\Eeq
where $d\sigma_n^2$ denotes a constant curvature space of dimension $n$
with a sectional curvature $K$, these equations reduce to
\begin{eqnarray}
&& H^2\equiv \left({\dot a\over a}\right)^2={2\kappa^2\over
n(n-1)}\rho -{K\over a^2},\\
&&{\cal D}\rho = - n h, \label {e:bg}\\
&&{\cal D}\rho_\alpha =- n (1-q_\alpha) h_\alpha,\label {e:bgc}
\end{eqnarray}
Here
\begin{equation}
{\cal D}:=a {d \over da},
\end{equation}
$h=\rho+P$ and $h_\alpha=\rho_\alpha+P_\alpha$ denote the entalpies of 
the total system and the $(\alpha)$-component, respectively, and
$q_\alpha$ is defined by
\Beq
q_\alpha={Q_\alpha\over nh_\alpha H};\quad
(Q_{(\alpha)\mu})=(Q_\alpha,0,\cdots,0).
\Eeq

 For scalar perturbations, the linear perturbation equations for
\Eq{EinsteinEq:general} are written in terms of gauge-invariant
quantities $\Delta$, $V$, $\Phi$, $\Psi$, $\Gamma$, and $\Pi$
representing the perturbation amplitudes of density, velocity,
curvature, gravitational potential, entropy and anisotropic stress,
respectively, as
\begin{eqnarray}
&&\D \Delta -n w \Delta = -C_K (1+w){k\over aH} V -(n-1)C_K w \Pi,\label{e:D}\\
&&\D V + V =
\nonumber \\
&& \quad 
{k\over aH} \Psi + {k\over aH} 
\left[{c^2_S \over 1+w}\Delta +{w \over 1+w}\Gamma \right]
-{n-1 \over n}{k\over aH} C_K {w \over 1+w}\Pi, \label{e:V}\\
&& \Delta = {2C_K\over nI_K}{k^2\over a^2H^2}\Phi,\\
&& (n-2)\Phi+\Psi=-\kappa^2{a^2\over k^2}p\Pi.
\end{eqnarray}
Here $w=P/\rho$, $c_s^2=\dot P/\dot\rho$ and 
\begin{equation}
I_K := 1+{K \over a^2 H^2}, \quad
C_K := 1-{n K \over k^2}.
\end{equation}
In applications it is often more convenient to rewrite these evolution 
equations in terms of $\Phi$ and $\Upsilon$ defined by
\begin{equation}
\Upsilon := {aH \over k}V,
\end{equation}
as
\begin{eqnarray}
&& {\cal D} \Phi + (n-2) \Phi =
 - {n \over 2} I_K (1+w) \Upsilon 
 - {n (n-1) \over 2} ({a H \over k})^2 I_K  w \Pi,
\label {e:Ds} \\
&& {\cal D} \Upsilon + {n \over 2} I_K (1+w) \Upsilon 
 - {K \over a^2 H^2} \Upsilon 
\nonumber \\
&& \quad
= 
 \Biggl\{ -(n-2)+ 
 {c^2_S \over 1+w} {2 \over n} ({k \over a H})^2 {C_K \over I_K} \Biggr\}
 \Phi 
 + {w \over 1+w} \Gamma
 \nonumber \\
&& \quad\quad+ \Biggl\{ - {n (n-1) \over 2} ({a H \over k})^2 w I_K 
  - {n-1 \over n} C_K {w \over 1+w} \Biggr\} \Pi.
\label {e:Vs}
\end{eqnarray}

On the other hand, from \Eq{EOM:general}, the perturbation equations
for each component are given by
\begin{eqnarray}
&& \D[\rho_\alpha \Delta_\alpha] + n\rho_\alpha \Delta_\alpha
= {n \over n-1}\kappa^2 h h_\alpha {a \over kH} (V -V_\alpha)
\nonumber \\
&& \qquad
- C_K \left[h_\alpha {k\over aH} V_\alpha 
 +(n-1)(1-q_\alpha)P_\alpha \Pi_\alpha\right] \nonumber \\
&& \qquad -n q_\alpha [P_\alpha \Gamma_\alpha+c^2_\alpha \rho_\alpha \Delta_\alpha]
+nh_\alpha q_\alpha E_\alpha 
+n (1-q_\alpha){aH \over k} h_\alpha F_\alpha, 
\label{e:Delp}\\
&& \D V_\alpha + V_\alpha = { k\over aH} \Psi + {k\over aH} 
\left({c^2_\alpha \over 1+w_\alpha}\Delta_\alpha 
+{w_\alpha \over 1+w_\alpha}\Gamma_\alpha \right)
\nonumber \\
&&  \qquad 
 -{n-1 \over n}{k\over aH} C_K {w_\alpha \over 1+w_\alpha}\Pi_\alpha
+F_\alpha.
\label{e:Vp}
\end{eqnarray}
Here $E_\alpha$ and $F_\alpha$ are the gauge-invariant quantities
representing the energy and the momentum transfer rate for the
component $(\alpha)$ at its own rest frame. These quantities are
related to the corresponding quantities at the rest frame of the total 
system, $E_{c\alpha}$ and $F_{c\alpha}$, by
\Beqr
&& E_{c\alpha}=E_\alpha+{a\dot Q_\alpha\over kQ_\alpha}(V_\alpha-V),\\
&& F_{c\alpha}=F_\alpha + nq_\alpha (V_\alpha-V).
\label{e:Fcalp}
\Eeqr
These quantities must satisfy the following constraint obtained from 
\Eq{EMTransferRate:constraint}:
\begin{equation}
 \sum_\alpha Q_\alpha E_{c \alpha}=0, \qquad
 \sum_\alpha h_\alpha F_{c \alpha}=0.
\label {e:EMconp}
\end{equation}

Eqs.(\ref {e:D}), (\ref {e:V}), (\ref {e:Delp}), (\ref {e:Vp}) and
(\ref {e:EMconp}) determine the dynamical behavior of the
multicomponent system completely.  However it is often more convenient
to use the variables representing the relative perturbations among
components defined by
\begin{eqnarray}
&&S_{\alpha \beta} = {\Delta_{c \alpha} \over 1+w_\alpha}
   -{\Delta_{c \beta} \over 1+w_\beta}, 
\label {e:relden}\\
&&V_{\alpha \beta}= V_\alpha -V_\beta,\\
&& \Gamma_{\alpha \beta} = {w_\alpha \over 1+w_\alpha}\Gamma_\alpha 
   -{w_\beta \over 1+w_\beta}\Gamma_\beta, \\
&&\Pi_{\alpha \beta}= {w_\alpha \over 1+w_\alpha}\Pi_\alpha 
   -{w_\beta \over 1+w_\beta}\Pi_\beta,\\
&&E_{\alpha \beta}= q_\alpha E_{c \alpha}-q_\beta E_{c \beta}, \\
&&F_{\alpha \beta}= F_{c \alpha}-F_{c \beta}.
\label {e:momdif}
\end{eqnarray}
The dynamical variables for each component are written 
in terms of these relative variables and those for the total system as
\begin{eqnarray}
&& {\Delta_{c \alpha} \over 1+w_\alpha}
={\Delta \over 1+w}+\sum_\gamma {h_\gamma \over h}S_{\alpha \gamma}, \\
&& V_\alpha = V +\sum_\gamma {h_\gamma \over h}V_{\alpha \gamma}, \\
&& {w_\alpha \over 1+w_\alpha}\Gamma_\alpha 
 =  {w \over 1+w}(\Gamma - \Gamma_{rel}) +
\sum_\gamma {h_\gamma \over h}\Gamma_{\alpha \gamma}, 
\label {e:gamcom}\\
&&{w_\alpha \over 1+w_\alpha}\Pi_\alpha 
 = {w \over 1+w}\Pi +
\sum_\gamma {h_\gamma \over h}\Pi_{\alpha \gamma},
\end{eqnarray}
where $\Gamma_\r{rel}$ is the contribution to $\Gamma$ 
from the relative entropy perturbations and given by
\Beqr
 P \Gamma_{rel} &=& {1 \over 2} \sum_{\alpha} \sum_{\beta}
 {h_{\alpha} h_{\beta} \over h} (1 - q_{\alpha}) (1 - q_{\beta})
 (c^2_\alpha - c^2_\beta) \times 
 \nonumber \\
 & & \Biggl[
 {\Delta_{c \alpha} \over (1+w_{\alpha}) (1-q_\alpha)}
 -{\Delta_{c \beta} \over (1+w_{\beta}) (1-q_\beta)}
          \Biggr]
 \nonumber \\
                &=& {1 \over 2} \sum_{\alpha} \sum_{\beta}
 {h_{\alpha} h_{\beta} \over h} 
 (c^2_\alpha - c^2_\beta) S_{\alpha \beta}
                 +
 {1 \over 1+w} {\Delta \over n H}
     \sum_\alpha c^2_\alpha Q_\alpha.
 \label {e:gamrel}
\Eeqr

The evolution equations for these relative variables are derived as
follows. First, by subtracting Eq.(\ref {e:V}) from Eq.(\ref {e:Vp}),
we obtain
\begin{eqnarray}
&& \D(V_\alpha -V) +(V_\alpha -V)
 ={k\over aH} (c^2_\alpha-c^2_S){\Delta \over 1+w}
 +{k\over aH} \sum_{\gamma} {h_\gamma \over h}c^2_\alpha S_{\alpha \gamma} 
 \nonumber \\
&& \qquad -{k\over aH}{w \over 1+w} \Gamma_{rel} 
 +{k\over aH} \sum_\gamma {h_\gamma \over h}[
       \Gamma_{\alpha \gamma} -{n-1 \over n}C_K \Pi_{\alpha \gamma}
    ]
 \nonumber \\
&&\qquad +F_\alpha +nc^2_\alpha(1-q_\alpha)(V_\alpha -V).
\label{e:difV}
\end{eqnarray}
Next from the relation  
\begin{equation}
{\Delta_{c \alpha}\over 1+w_\alpha}
={\Delta_{\alpha}\over 1+w_\alpha}
 +{1 \over 1+w_\alpha} {\dot \rho_\alpha  \over \rho_\alpha}
  {a \over k}(V_\alpha-V),
\end{equation}
Eq.(\ref{e:Delp}), and Eq(\ref{e:difV}), it follows that
\begin{eqnarray}
&&
\D({\Delta_{c \alpha}\over 1+w_\alpha})
 = -{k\over aH} V_\alpha +{n K \over kaH}V 
   +n\{
     (1-q_\alpha)c^2_S - q_\alpha(1+c^2_\alpha)          
     \} {\Delta \over 1+w}
 \nonumber \\
&&  \qquad
+n(1-q_\alpha){w \over 1+w}\Gamma
   -nq_\alpha(1+c^2_\alpha) \sum_{\gamma}{h_\gamma \over h}S_{\alpha \gamma}
 \nonumber \\
&& \qquad
   -n{w_\alpha \over 1+w_\alpha}\Gamma_\alpha
   -(n-1)C_K(1-q_\alpha){w \over 1+w}\Pi
   +nq_\alpha E_{c \alpha}.
\label{e:Delcp}
\end{eqnarray}
 From these equations we obtain the following evolution equations for
$S_{\alpha \beta}$ and $V_{\alpha \beta}$:
\begin{eqnarray}
&& \D S_{\alpha \beta}
 +{n \over 2}[
    q_\alpha (1+c^2_{\alpha})+q_\beta (1+c^2_{\beta})
    ] 
    S_{\alpha \beta}
 \nonumber \\ 
 && \quad +{n \over 2}[
    q_\alpha (1+c^2_{\alpha})-q_\beta (1+c^2_{\beta})
    ] 
    \sum_{\gamma} {h_\gamma \over h} 
     (S_{\alpha \gamma}+S_{\beta \gamma})
 \nonumber \\
&& \quad=-{k\over aH} V_{\alpha \beta}
   -n \Gamma_{\alpha \beta} +n E_{\alpha \beta}
 \nonumber \\
&&\qquad   -n[(q_\alpha-q_\beta) c^2_S +q_\alpha(1+c^2_\alpha)
              -q_\beta(1+c^2_\beta)]{\Delta \over 1+w}
 \nonumber \\
&&\qquad -n(q_\alpha-q_\beta){w \over 1+w}\Gamma
   +(n-1)C_K (q_\alpha-q_\beta){w \over 1+w}\Pi,
\label{e:ED}\\
&& \D V_{\alpha \beta}+ V_{\alpha \beta}
   -{n \over 2} [
     c^2_\alpha +c^2_\beta -q_\alpha(1+c^2_\alpha)
                           -q_\beta(1+c^2_\beta)
    ] V_{\alpha \beta}
 \nonumber \\
&& \qquad -{n \over 2} [
     c^2_\alpha -c^2_\beta -q_\alpha(1+c^2_\alpha)
                           +q_\beta(1+c^2_\beta)
    ] \sum_\gamma {h_\gamma \over h}
      (V_{\alpha \gamma}+V_{\beta \gamma})
 \nonumber \\
&& \quad   = {k\over aH} (c^2_\alpha -c^2_\beta){\Delta \over 1+w}
   + {k \over 2aH} (c^2_\alpha +c^2_\beta) S_{\alpha \beta}
   + {k \over 2aH}(c^2_\alpha -c^2_\beta)
     \sum_{\gamma} {h_\gamma \over h} 
      (S_{\alpha \gamma}+S_{\beta \gamma})
 \nonumber \\
&& \qquad   +{k\over aH} \Gamma_{\alpha \beta}
   -{n-1 \over n}{k\over aH} C_K \Pi_{\alpha \beta}
   +F_{\alpha \beta}.
\label{e:EV}
\end{eqnarray}

\section{
Upper bound on the growth rate of solutions
               to a first-order differential equation system
        }

In this appendix we explain a general technique to evaluate an upper
bound on the norm of the solution \bm{X} to the first differential
equation system
\begin{equation}
{d \over dt}\bm{X}= \bg{\Omega}(t)\bm{X}+\bm{S}(t)
    .
\end{equation}
Here $\bm{X}$ and $\bm{S}$ are $n$ column vectors,
and $\bg{\Omega}$ is an $n \times n$ matrix. 
If we define the norm of the solution $\bm{X}$ by
\begin{equation}
\| \bm{X} \| ^2 := \bm{X}^\dagger \bm{X} 
    ,
\end{equation}
it obeys the equation 
\begin{equation}
{d \over dt} (\| \bm{X} \| ^2) 
  =  \bm{X}^\dagger \bg{\Omega}_H \bm{X}
    + \bm{S}^\dagger \bm{X} +  \bm{X}^\dagger \bm{S}
    ,
\label {e:sqr}
\end{equation}
where $\bg{\Omega}_H$ is an hermitian matrix defined by 
\begin{equation}
\bg{\Omega}_H :=  \bg{\Omega}^\dagger + \bg{\Omega}
    .
\end{equation}

The least upper bound of the right-hand side of this equation is
determined by the maximum eigenvalue of $\bg{\Omega}_H$.  However, in
many cases this simple method is not practical because it gives a
complicated expression.  In such cases it is useful to decompose
$\bg{\Omega}_H$ into a sum of hermitian matrices $\bg{\Omega}_{Hi}$,
($1\leq i \leq k$) as
\begin{equation}
\bg{\Omega}_H = \sum_{i=1}^k \bg{\Omega}_{Hi}
    ,
\label {e:decom}
\end{equation}
so that the maximum eigenvalue of each $\bg{\Omega}_{Hi}$ has a
simple expression.  Since the hermitian matrix $\bg{\Omega}_{Hi}$ can
be diagonalized by some unitary matrix and its eigenvalues are real,
we obtain
\begin{equation}
\bm{X}^\dagger \bg{\Omega}_{Hi} \bm{X}
   \leq \lambda_{mi} \bm{X}^\dagger \bm{X}
    ,
\label {e:prop1}
\end{equation}
where $\lambda_{mi}$ is the largest eigenvalue of the hermitian matrix
$\bg{\Omega}_{Hi}$.  Hence by applying the Cauchy-Schwartz inequality
\begin{equation}
\Tr \bm{A}^\dagger \bm{B}
\leq
 (\Tr \bm{A}^\dagger \bm{A})^{1 / 2}
  (\Tr \bm{B}^\dagger \bm{B})^{1 / 2}
\label {e:prop2}
\end{equation}
to Eq.({\ref {e:sqr}}), we obtain
\begin{equation}
{d \over dt}(\| \bm{X} \|)
  - {\lambda_m \over 2} \| \bm{X} \|
    - \| \bm{S} \|
\leq 0
    ,
\label {e:difineq}
\end{equation}
where
\begin{equation}
\lambda_m := 
  \sum_{i=1}^k \lambda_{mi} 
     .
\end{equation}
By integrating the inequality ({\ref {e:difineq}}),
we obtain
\begin{equation}
\| \bm{X}(t) \| 
  \leq 
    \exp \left( \int_0^t 
    {\lambda_m (t^\prime) \over 2} dt^\prime \right) 
        \| \bm{X}(0) \|
    + \int_0^t dt^\prime 
         \exp \left( \int_{t^\prime}^t 
    {\lambda_m (t^{\prime \prime})\over 2} dt^{\prime \prime} 
              \right)
            \| \bm{S}({t^\prime}) \|
     .
\label {e:uplim}
\end{equation}
Notice that the right-hand side of the inequality ({\ref {e:uplim}})
is a monotonically increasing functional of $\lambda_m (s)$, $(0 \leq
s \leq t)$.  Therefore, even if we replace $\lambda_m$ by a larger
value, the inequality ({\ref {e:uplim}}) still holds.

In the case $\lambda_m (t) < 0$ and $| \lambda_m (t) | \gg 1$, we can
derive a simpler estimate on $\| \bm{X} \|$.  To see this, let us
introduce $x(t')$ defined by
\Beq
x(t'):=-\int_{t'}^t \lambda_m(t'')dt''.
\Eeq
Then the second term in the right-hand side of Eq.({\ref {e:uplim}} ) 
is written as
\Beq
\int_0^tdt'\exp\left({1\over2}\int_{t'}^t\lambda_m(t'')dt''\right)
\|\bm{S}(t')\|=\int_0^\Lambda dx\exp\left(-{1\over2}x\right)
{ \|\bm{S}\|_x \over |\lambda_m|_x } ,
\Eeq
where
\Beq
\Lambda=\int_0^t|\lambda_m(t')|dt'.
\Eeq
Hence, if $ \|\bm{S}\| / |\lambda_m| $ is a slowly varying function 
of $t$ and $\Lambda\gg1$, we obtain
\begin{equation}
 \int_0^t dt^\prime 
         \exp ( \int_{t^\prime}^t 
     {\lambda_m (t^{\prime \prime}) \over 2} dt^{\prime \prime} 
              )
            \| \bm{S}({t^\prime}) \|
 \leq
 \Order{ \| \bm{S}(t) \| \over |\lambda_m (t)| }.
\label {e:uplimp}
\end{equation}
%


\end{document}